%% file: Moriond.tex
\documentclass[11pt]{article}
\usepackage{moriond,epsfig}
\usepackage{psfig,graphicx,changebar}
\usepackage{amsmath}
\include{newcom}

\def\Journal#1#2#3#4{{#1} {\bf #2}, #3 (#4)}

\def\NPB{{\em Nucl. Phys.} B}
\def\PLB{{\em Phys. Lett.}  B}

\def\PR{\em Phys. Rev.}

\def\NAT{\em Nature}
\def\EPJC{{\em Eur. Phys. J.} C}
\def\CPC{\em Computer Physics Commun.}

\def\be{\begin{equation}}
\def\ee{\end{equation}}
\def\bea{\begin{eqnarray}}
\def\eea{\end{eqnarray}}

\begin{document}
\vspace*{4cm}
\title{Bose-Einstein correlations in WW events at LEP}

\author{ Nick van Remortel}

\address{Department of Physics, Universiteit Antwerpen \\ 
Universiteitsplein 1,
B-2610 Antwerpen \\
email: Nick.van.Remortel@cern.ch}

\maketitle\abstracts{
The current status of the LEP results on Bose-Einstein correlations is discussed.
Emphasis is given to the measurement of Bose-Einstein
correlations between decay products from different W's, in an energy
range between 172 and 209 GeV, dependent on the experiment.
For the first time all four LEP experiments conclude
 that no evidence for correlations between pions from different W's
is seen at the current level of precision.}

\section{Introduction}
  Correlations between pairs/multiplets of identical bosons (in the
  simplified experimental practice, all like-sign particles are considered
  instead) are a well known phenomenon, yet the understanding of the effect
  is far from complete. 
  Let us start with the original observation of the effect by Goldhaber and
  collaborators \cite{goldhab}.
  In order to give an interpretation to their observations the authors started
  from the assumption that contributions from different bosons to the
  measured intensity add  incoherently. A strict analogy with the  Hanbury-Brown-Twiss effect \cite{HBT}
 in astronomy was found. In both cases the measured intensity interference
  reflects the geometry of the emitter. However, using this scenario too
  strictly one soon runs into interpretational problems. The shape of the
  correlation functions measured since Goldhaber's observation
  until the LEP measurements does not reflect the size of the freeze-out volume
  at the moment of hadronisation which extends up to several fermi.
  Most observations, except in heavy ion collisions, indicate source sizes of
  the order of one fermi.

  An alternative model was
  proposed by Andersson and Ringn\'{e}r \cite{lund}, in which
  the correlations appear as a coherent effect related to the
  symmetrisation of the quantum-mechanical amplitude corresponding to the
  full process of particle production in the fragmentation of the Lund
  string. The strong point here is that the introduction of the Bose-Einstein
  effect becomes less arbitrary than before. It essentially depends  on two
  fundamental parameters of the Lund model, the string tension $\kappa$ and
  the hadronisation cutoff $b$ parameter. In this case one can obtain source
  sizes compatible with experimental observations. This model has however
  one fundamental restriction: only bosons from the same string can be
  subjected to the Bose-Einstein effect, provided that there is no Colour
  Reconnection at parton level.

    For a simple hadronic system like $q\bar{q}$   from a $Z^0$ decay, it
  may be impossible to decide between the two possibilities, since the 
  incoherent approach leaves a freedom of the choice of the input
  particle density, which can be adjusted to reproduce the observed data.

    The study of correlations between two close hadronic systems,
  such as hadronically decaying pairs of WW/ZZ bosons, can eventually
  help to distinguish between the two possibilities. In the incoherent
  scenario, the difference between correlations
  within a single hadronic system, and correlations between the two systems, 
  should depend only on the overlap of the two systems (sources).  In the
  coherent scenario, the correlations between the two systems may 
  not exist at all, even for overlapping sources (as long as there is no
  interaction -colour flow- between these).  
 
    The measurement of inter-W correlations is also important for
  the estimate of the systematic bias in the measurement of the W mass
  via the direct reconstruction of measured decay products. 
  A better understanding of the physical origin of the observed correlations
  is however necessary to ensure a reliable prediction for
  the uncertainty on the W mass measurement.
\section{Analysis methods}
It is common practice to investigate BEC between particles coming
from different W's by means of a two-particle correlation function in terms
of the Lorenz-invariant four-momentum transfer $Q=\sqrt{-(p_1-p_2)^2}$:
\begin{equation}
\label{eq:01}
R(Q)=\frac{\rho(Q)}{\rho_0(Q)},
\end{equation}
where $\rho_0(Q)$ represents the two-particle density without the
Bose-Einstein effect.
This density is non-existent in nature and is known as the so-called
normalization or reference sample problem.
We will see that many experiments address this problem in different ways,
each with their own degree of model and detector dependence.
A widely used implementation of the Bose-Einstein correlation effect in Monte
Carlo generators is the LUBOEI \cite{luboei} code, included in JETSET
\cite{jetset}. 
Experiments use different versions of this code and tune the Monte Carlo samples to their
$Z^0$ data.
\subsection{OPAL analysis}
The OPAL collaboration has published an analysis \cite{opal} for a total collected
statistics of 250 \pbi.
In this analysis the two-particle correlation function is constructed
using unlike-sign pairs as a reference sample and making a double ratio with
the correlation function obtained for a Monte-Carlo sample without
Bose-Einstein correlations at all:
\begin{equation}
\label{eq:02}
C(Q)=\frac{N^{data}_{\pm \pm}(Q)}{N^{data}_{+-}(Q)}/\frac{N^{MC}_{\pm \pm}(Q)}{N^{MC}_{+-}(Q)}.
\end{equation}
This is done for three samples:
fully hadronic WW decays, semi-leptonic WW decays and $q\bar{q}$ events
selected as fully-hadronic WW events.
One can assume that each of these 3 correlation functions can be written as the sum
of 3 independent and more interesting correlation functions. For example one
can write the correlation function for the fully hadronic sample as
\begin{equation}
\label{eq:03}
C^{had}(Q) = P^{s}_{had}(Q)C^{s}(Q)+
  P^{Z^*}_{had}(Q)C^{Z^*}_{bg}(Q) +(1-P^{s}_{had}(Q)-P^{Z^*}_{had}(Q))C^{d}(Q), 
\end{equation}
where \text{$C^{s}(Q)$, $C^{Z^*}_{bg}(Q)$, $C^{d}(Q)$} represent the correlation
functions for particle pairs originating from the same W, the $Z^0$ background and
for pairs originating from different W's, each with their own probabilities
$P(Q)$, obtained from MC samples without Bose-Einstein Correlations.
In a next step OPAL makes a simultaneous fit to the three measured
correlation function using the expression:
\begin{equation}
\label{eq:04}
C^{s,d,Z}(Q)=N(1+f_{\pi}(Q)\lambda^{s,d,Z} e^{-R^2 Q^2}),
\end{equation}
where \text{$f_{\pi}(Q)$} is the probability that a given particle pair is indeed a
pair of pions, obtained from Monte Carlo. 
Taking into account the distance between the W decay vertices one can impose
a constraint on the radii:
\begin{equation}
\label{eq:05}
(R^d)^2=(R^s)^2+(correction)^2. 
\end{equation}
This gives a fit result of
\begin{equation*}
\begin{split}
& \lambda^s=0.69\pm 0.12(stat) \pm 0.06(syst), \\ &\lambda^d=0.05\pm 0.67(stat) \pm 0.35(syst),
\end{split}
\end{equation*}
leading to the conclusion that with this method and at the current level of
precision it is impossible to establish whether BEC between different W's
exists or not.
\subsection{ALEPH analysis}
\begin{figure}[t]
\begin{minipage}{.46 \linewidth}
    \centering\epsfig{figure=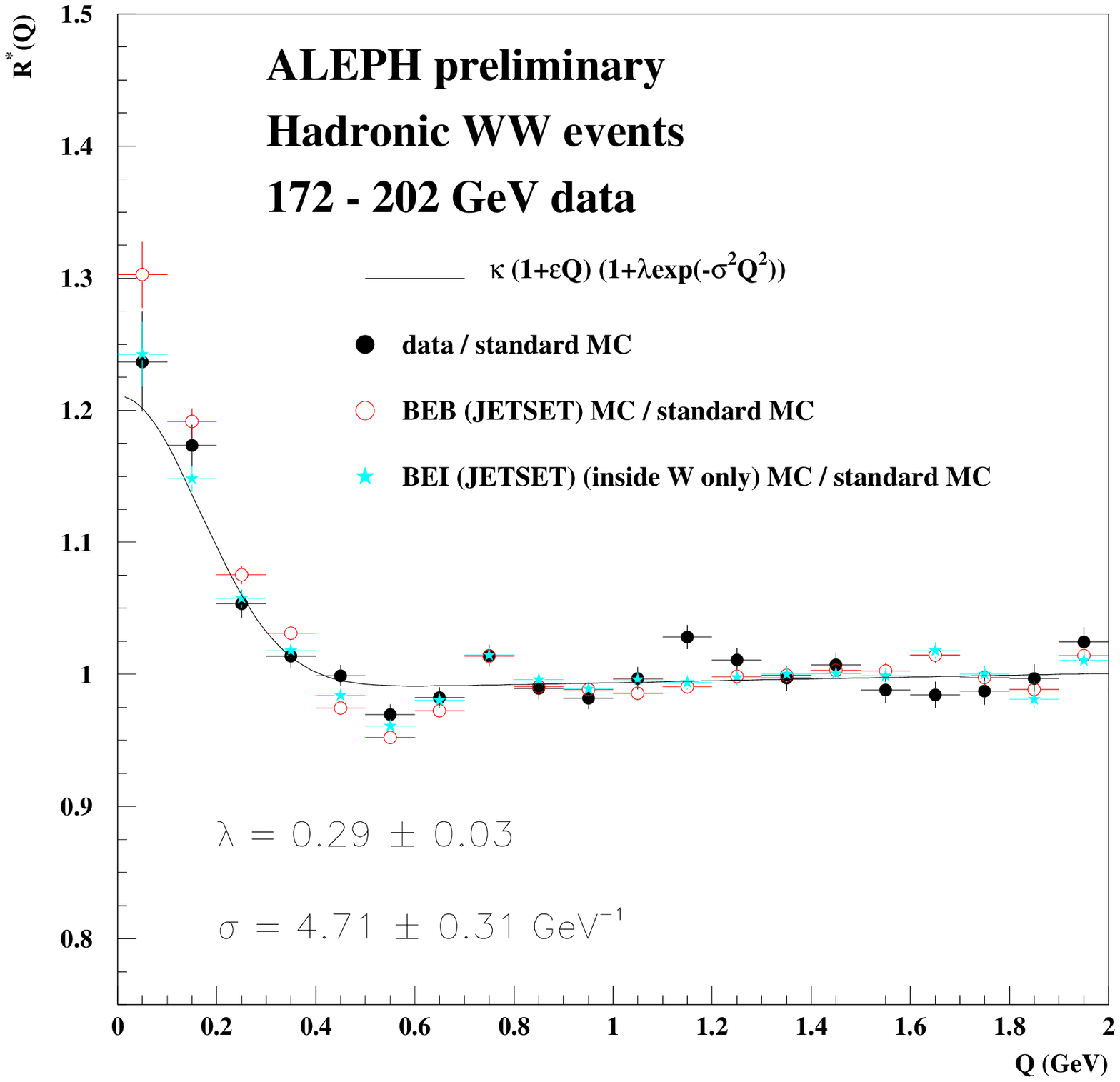,width=6.3cm}
    \caption{The $R^*(Q)$ distribution for data compared with BE3 model
    predictions. Only statistical errors are shown. The solid curve shows the
    fit result to the data.} \label{figaleph1}
\end{minipage}\hfill
\begin{minipage}{.46 \linewidth}
    \centering\epsfig{figure=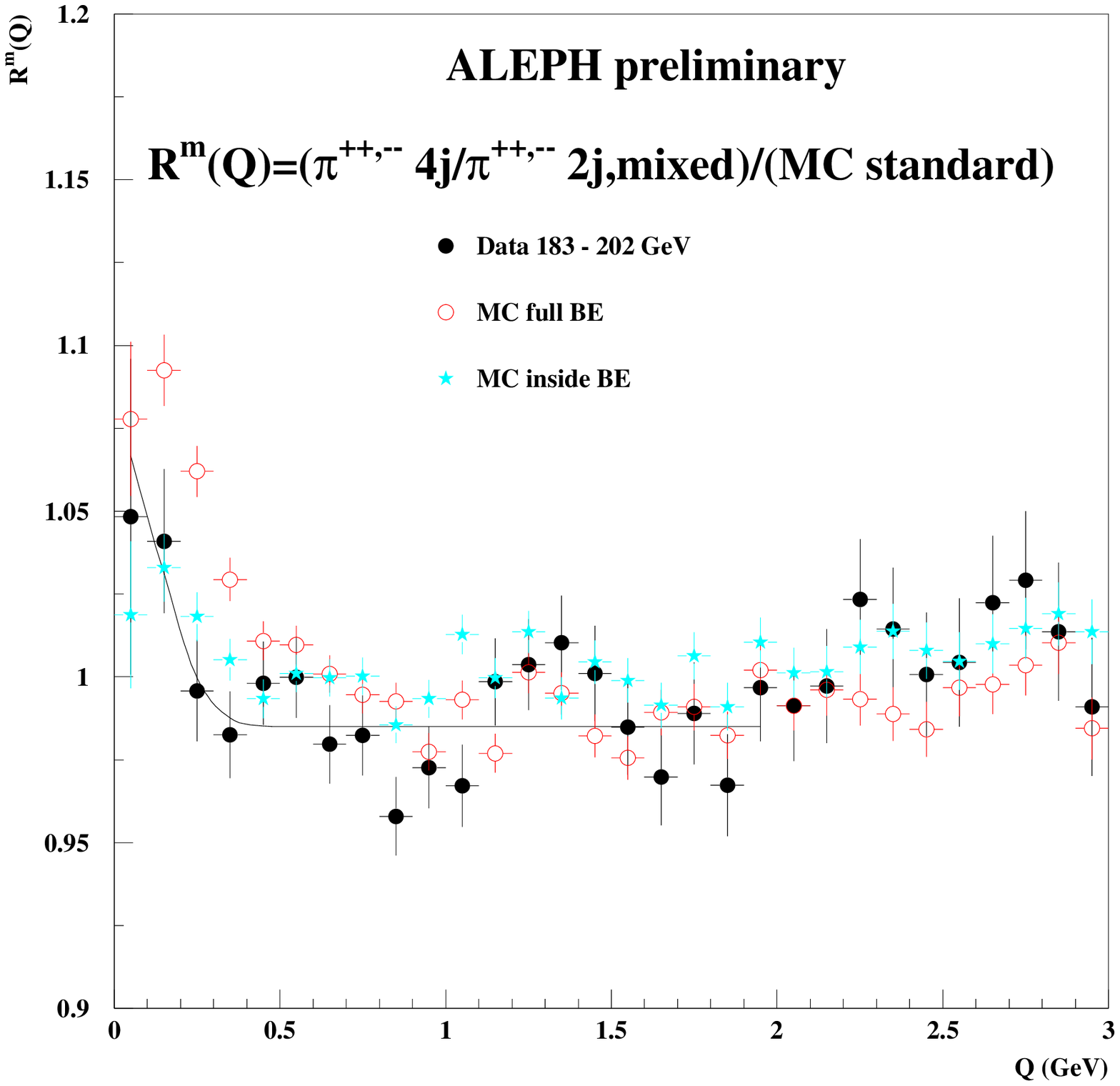,width=6.cm}
    \caption{The $R^m(Q)$ distribution for data compared with model
    predictions. The solid curve is the fit result to the data.} \label{figaleph2}
\end{minipage}
\end{figure}

The ALEPH collaboration has published results \cite{aleph1} for the energy range
between 172 and 189 \GeV. An update including energies up to
202 GeV was submitted to ICHEP2000 \cite{aleph2}.
Similar to OPAL, ALEPH also uses unlike-sign pairs as a reference sample and
corrects for resonance decays and detector effects by making a double ratio
with a MC sample without BEC at all. Since the $q\bar{q}$ background might fake a
possible inter-W BEC signal it was decided to add the background fraction to
the MC reference without BEC included. In this way the two-particle
correlation function becomes
\begin{equation}
\label{eq:06}
R^*(Q)=
\frac{N^{data}_{\pm \pm}(Q)}{N^{data}_{+-}(Q)}/\frac{N^{MC(WW+q\bar{q})}_{\pm
    \pm}(Q)}{N^{MC(WW+q\bar{q})}_{+-}(Q)} 
\end{equation}
The distribution of $R^*(Q)$ is compared between data and two Bose-Einstein
models based on the LUBOEI BE3 algorithm, tuned on $Z^0$ data, as can be seen
in Fig.~\ref{figaleph1}.
Fits to this distribution for data and models are made using expression
\begin{equation}
\label{eq:07}
R^*(Q)=\kappa (1+\epsilon Q)(1+\lambda e^{-\sigma^{2}Q^2}).
\end{equation}
The results of the fits are compared by integrating over the correlation
signal
\begin{equation}
\label{eq:08}
I = \int \limits_0^\infty\lambda e^{-\sigma ^{2}Q^2}dQ = \frac{\sqrt{\pi}}{2}\frac{\lambda}{\sigma}.
\end{equation}
For this measurement ALEPH finds that the value of $I$ for the data  is
compatible with the value of $I$ for the BE3 model in which only intra-W BEC
are present. The BE3 model with intra+inter BEC is disfavored at the level of
2.2 $\sigma$.

In a second method mixed semi-leptonic events are used as reference
sample. Again a double ratio with a MC sample without BEC including the
$q\bar{q}$ background is used, and the two-particle correlation function
becomes:
\begin{equation}
\label{eq:09}
R^m(Q) =
\frac{N^{4q data}_{\pm \pm}(Q)}{N^{mixed}_{\pm \pm}(Q)}/\frac{N^{MC(4q+q\bar{q})}_{\pm
    \pm}(Q)}{N^{MC(mixed)}_{\pm \pm}(Q)}. 
\end{equation}
This distribution (see Fig.~\ref{figaleph2}) is again fitted with a gaussian parametrisation and
integrals are compared. In this case the inter+intra BEC scenario is
disfavored at the level of 3.1 \sig (stat only).\\
\subsection{L3 analysis}
\begin{figure}[t]
\begin{minipage}{.46 \linewidth}
    \centering\epsfig{figure=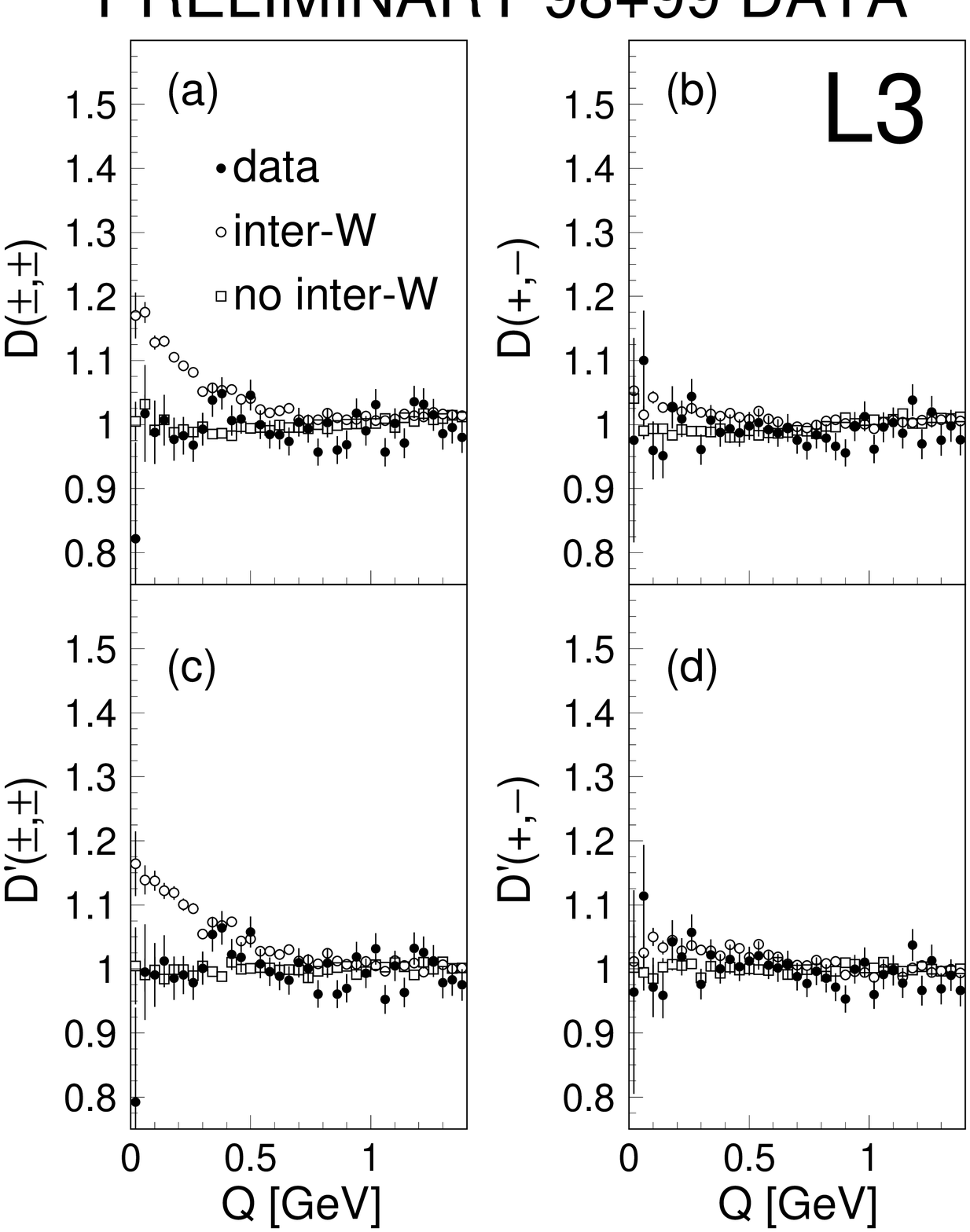,width=8.3cm}
    \caption{The $D$ and $D'$ distribution for data compared with BE32 model
    predictions. Only statistical errors are shown. Bin-to-bin correlations
    are not considered} \label{l3}
\end{minipage}\hfill
\begin{minipage}{.46 \linewidth}
    \centering\epsfig{figure=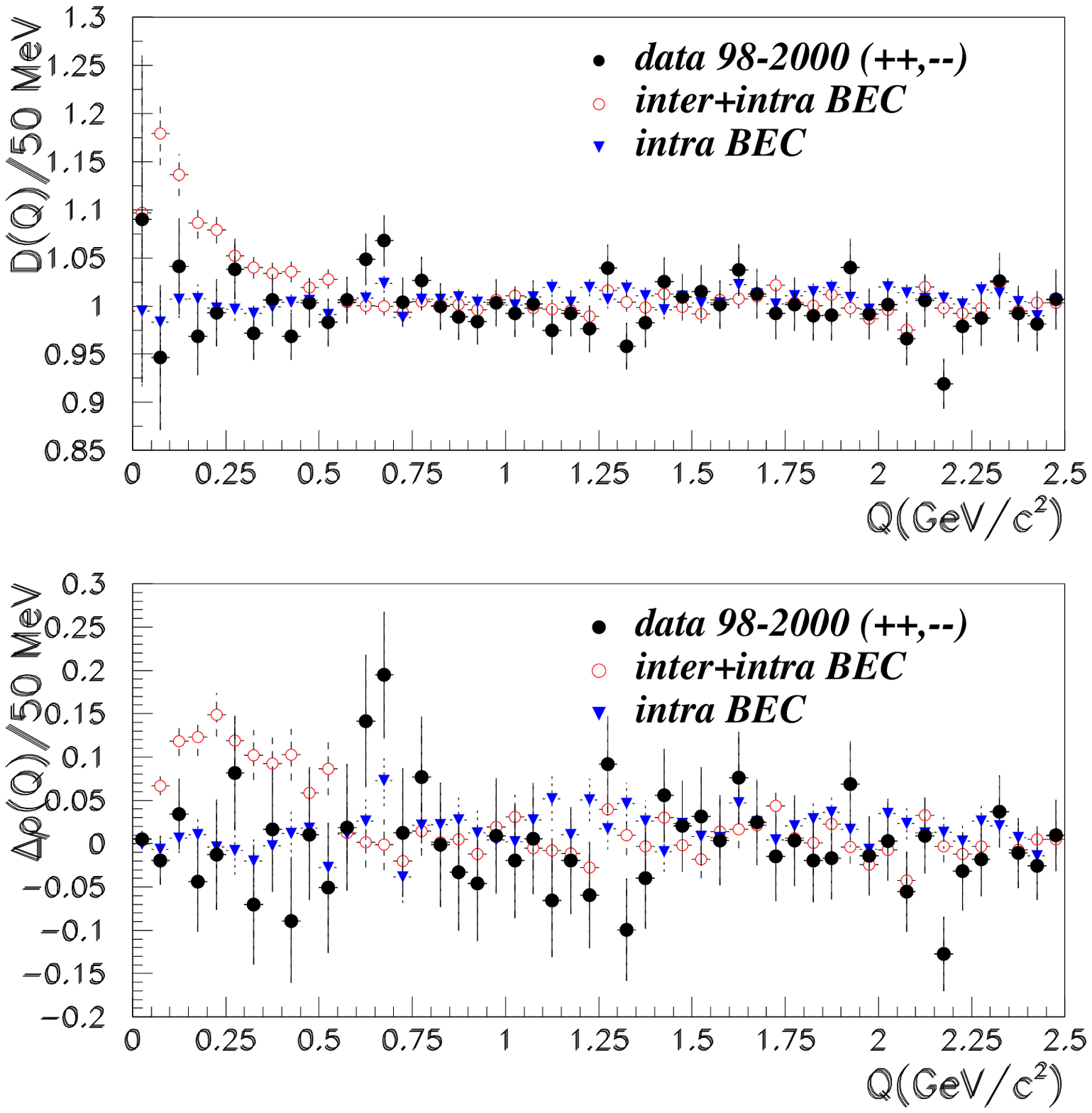,width=8.cm}
    \caption{The $D$ and $\Delta \rho$ distribution for the DELPHI data
    compared with BE32 model predictions. Only statistical errors are
    shown. Bin-to-bin correlations are not considered.} \label{delphi}
\end{minipage}
\end{figure}

The L3 analysis relies on a rigorous mathematical treatment \cite{kit} and
generalizations thereof \cite{edi} and is published \cite{l3} for the
collected data at 189 GeV. A new update has been given for this conference,
including the 192-202~GeV data.
In their formalism one can write the two-particle densities for independently
decaying W's as
\begin{equation}
\label{eq:10}
\rho^{WW}(1,2)=2\rho^W(1,2)+2\rho^{WW}_{mix},
\end{equation}
where the second term on the right-hand side of Eq.~\ref{eq:10} is obtained by
mixing 2 semi-leptonic events.
In the absence of inter-W BEC the ratio of the left-hand side and right-hand
side of Eq.~\ref{eq:10}, which is called $D$, should be compatible with one.
After subtracting 18.6\% $q\bar{q}$ background from the fully hadronic term
$\rho^{WW}(1,2)$, using the LUBOEI BE0 model, L3 makes a double ratio by
dividing the $D$ distribution for the data by the same distribution obtained
with a MC sample without any BEC included. This variable is called $D'$ and is
fitted with a gaussian expression. Both distributions are shown in Fig.~\ref{l3}.
The fitted value for the correlation strength $\Lambda$ is compatible with zero:
\begin{equation*}
\Lambda = 0.013 \pm 0.018(stat) \pm 0.015(syst).
\end{equation*}
Comparison with the inter+intra BEC BE32 model tuned at the $Z^0$ data gives
a deviation from the data of 4.7 $\sigma$.\\
\subsection{DELPHI analysis}
The DELPHI analysis has been updated ~\cite{dnote} for a total collected
statistics of $531$~pb${}^{-1}$ including energies from 189-209 \GeV.
DELPHI used the same formalism as L3 and also studies the
difference between the right-hand side and left-hand side of Eq.~\ref{eq:10},
which is called $\Delta \rho (Q)$. This distribution, together with the $D$
distribution, is shown in Fig.~\ref{delphi}. The $q\bar{q}$ background
contamination in the data amounts to 15\% and is subtracted using the BE32 model
tuned on $Z^0$ data. Since this background subtraction is a delicate point,
DELPHI tried to investigate how well the BE32 model describes the $q\bar{q}$
events which are selected as WW events. This was done using 4 jet $Z^0$
events and high energy $q\bar{q}$ events with an anti-WW tag. The largest
disagreement between the model and data did not exceed the 10\% level. This
study is still ongoing. In order to stay as model independent as possible
DELPHI does not construct a $D'$ distribution and makes a fit directly to the $D$
variable with the following expression:
\begin{equation}
D(Q)=N (1+\epsilon Q)(1+\Lambda e^{-\sigma Q})
\end{equation}
After fixing $\sigma$ to 1.01 fm, as was fitted for the inter+intra
BEC model prediction, DELPHI finds a value of $\Lambda^{dat}$ compatible with
zero.
\begin{equation*}
\Lambda^{dat}= -0.038 \pm 0.057(stat) \pm 0.06(syst)
\end{equation*}
The systematic error is still under study and contains for the moment only
the contributions from the background subtraction (0.05) and from the mixing
method (0.03). However, it is assumed that these two contributions are the
dominant ones.
When comparing the fitted value of $\Lambda$ of the data with the inter+intra
BEC BE32 model prediction, DELPHI disfavors the model at the level of 3.2
$\sigma$. 
\section{Summary}
It is important to note that for the first time the 4 LEP experiments obtain
consistent conclusions. The LUBOEI models tuned on the $Z^0$ data
from each experiment, and  which include BEC between different W's, are excluded by
all experiments with varying significance. The LEP experiments are on the
way to converge on measurement techniques as proposed in \cite{kit,edi} ,which is very
promising.
It is my question to the W-mass measurement community whether they will still use these models to estimate
their systematic errors. What is clear for me is that WW events will not tell
us much more about the ongoing discussion on incoherence and coherence, and the
easy but rather restrictive variable $Q$ might not be the ideal one to be used.
Certainly a study of multi-string events from LEP1 would be very interesting
to address this problem \cite{edi}.

\end{document}

%% file: newcom.tex
\def\leqsim{\mathbin{\;\raise1pt\hbox{$<$}\kern-8pt\lower3pt\hbox{$\sim$}\;}}
\def\geqsim{\mathbin{\;\raise1pt\hbox{$>$}\kern-8pt\lower3pt\hbox{$\sim$}\;}}

\def\p#1{\mbox{$ \mbox{\bf p}_1                                         $}}

\newcommand{\ee}      {\mbox{$ {\, \mathrm e}^+ {\mathrm e}^-                 $}}

\newcommand{\GeV}     {\mbox{$ {\mathrm{GeV}}                              $}}

\newcommand{\pbi}     {\mbox{$ {\mathrm{pb}}^{-1}                          $}}

\newcommand{\sig}      {\mbox{$ {\mathrm \sigma}  \quad            $}}

\def\NPB#1#2#3{{\rm Nucl.~Phys.} {\bf{B#1}} (19#2) #3}
\def\PLB#1#2#3{{\rm Phys.~Lett.} {\bf{B#1}} (19#2) #3}

\def\PR#1#2#3{{\rm Phys.~Rep.} {\bf#1} (19#2) #3}

\def\CPC#1#2#3{{\rm Comp.~Phys.~Comm.} {\bf#1} (19#2) #3}

\def    \missEt      {\ifmmode{/\mkern-11mu E_t}\else{${/\mkern-11mu E_t}$}\fi}
\def    \missE       {\ifmmode{/\mkern-11mu E}\else{${/\mkern-11mu E}$}\fi}
\def    \missp       {\ifmmode{/\mkern-11mu p}\else{${/\mkern-11mu p}$}\fi}
\def    \misspt      {\ifmmode{/\mkern-11mu p_t}\else{${/\mkern-11mu p_t}$}\fi}